\documentclass[12pt]{article} 
\RequirePackage[colorlinks,citecolor=blue,urlcolor=blue]{hyperref}
\usepackage[utf8]{inputenc} 
\usepackage{hyperref}
\input {mathMacros.tex}
\usepackage{mathtools}
\usepackage{geometry} 
\usepackage{graphicx} 

\usepackage{booktabs} 
\usepackage{array} 
\usepackage{paralist} 
\usepackage{verbatim} 
\usepackage{subfig} 
\usepackage{amsbsy}
\usepackage{amsmath}
\usepackage{amssymb}
\usepackage{amsthm}
\usepackage{cite}
\usepackage{float}
\usepackage[toc,page]{appendix}
\usepackage{natbib}
\usepackage{lineno}
\usepackage{mathrsfs}
\usepackage{lineno}
\usepackage{todonotes}
\usepackage{setspace}
\usepackage{natbib}
\usepackage{multirow}
\usepackage{xargs}
\usepackage{fullpage}
\usepackage{bbold}

\allowdisplaybreaks

\usepackage{fancyhdr} 
\usepackage{sectsty}
\allsectionsfont{\sffamily\mdseries\upshape} 

\usepackage[nottoc,notlof,notlot]{tocbibind} 
\usepackage[titles,subfigure]{tocloft} 

\DeclarePairedDelimiter\set\{\}
\usepackage{fancyhdr} 
\usepackage{sectsty}
\allsectionsfont{\sffamily\mdseries\upshape} 

\usepackage[nottoc,notlof,notlot]{tocbibind} 
\usepackage[titles,subfigure]{tocloft} 

\begin{document}
\begin{flushright}
Version dated: \today
\end{flushright}

\bigskip
\medskip
\begin{center}

\noindent{\Large \bf Shrinkage-based random local clocks with scalable inference}

\noindent{\normalsize \sc
	Alexander A. Fisher$^{1}$\\
	Xiang Ji$^{2}$ \\
	Akihiko Nishimura$^{3}$ \\
	Philippe Lemey$^{4}$ \\
	    and
    Marc A.~Suchard$^{1,5,6}$}\\
\noindent {\small
	\it $^1$Department of Biomathematics, David Geffen School of Medicine at UCLA, University of California,
	Los Angeles, United States \\
	\it $^2$Department of Mathematics, School of Science \& Engineering, Tulane University, United States \\
	\it $^3$Johns Hopkins University, Baltimore, MD, United States \\
	\it $^4$Department of Microbiology, Immunology and Transplantation, Rega Institute, KU Leuven, Leuven, Belgium \\
  \it $^5$Department of Biostatistics, Jonathan and Karin Fielding School of Public Health, University
  of California, Los Angeles, United States} \\
  \it $^6$Department of Human Genetics, David Geffen School of Medicine at UCLA, Universtiy of California,
  Los Angeles, United States

\end{center}
\medskip
\noindent{\bf Corresponding author:} Marc A. Suchard, Departments of Biostatistics, Biomathematics, and Human Genetics,
University of California, Los Angeles, 695 Charles E. Young Dr., South,
Los Angeles, CA 90095-7088, USA; E-mail: \url{msuchard@ucla.edu}\\

\vspace{1in}

\clearpage

\begin{abstract}
	Local clock models propose that the rate of molecular evolution is constant within phylogenetic sub-trees.
	Current local clock inference procedures
	scale poorly to large taxa problems, impose model misspecification, or require a priori knowledge of the existence and location of clocks.
	To overcome these challenges, we present an autocorrelated, Bayesian model of heritable clock rate evolution that leverages heavy-tailed priors with mean zero to shrink  increments of change between branch-specific clocks.
	We further develop an efficient Hamiltonian Monte Carlo sampler that exploits closed form gradient computations to scale our model to large trees.
	Inference under our shrinkage-clock exhibits an over 3-fold speed increase compared to the popular random local clock when estimating branch-specific clock rates on a simulated dataset.
	We further show our shrinkage-clock recovers known local clocks within a rodent and mammalian phylogeny.
	Finally, in a problem that once appeared computationally impractical, we investigate the heritable clock structure of various surface glycoproteins of influenza A virus in the absence of prior knowledge about clock placement.



\end{abstract}

\section{Introduction}


Molecular clock models are ubiquitous phylogenetic instruments for divergence-time estimation with applications ranging from timing placental mammal radiation \citep{springer2003placental} to estimating influenza diversity \citep{davidson2014molecular}.
To capture clock rate variation along the lineages of a phylogeny,
\cite{thorne1998estimating} propose an autocorrelated, or ``heritable" rate model while others \citep{yoder2000estimation,drummond2010bayesian} assume there exist, at most, a small number of ``local" clocks on any given tree.
In each case, closely related lineages maintain similar or even identical evolutionary rates.
Autocorrelated rate models are computationally appealing due to the induced smooth transition in rate from parent to child node along the tree but may inappropriately shrink large rate changes between adjacent nodes \citep{smith2010uncorrelated}.
On the other hand, local clock models allow large rate changes to exist but can be computationally unpalatable on large problems due to the combinatorial complexity of choosing (or learning) the number and location of local clocks.
When these quantities are simultaneously learned with the tree, \citet{drummond2010bayesian} call this the random local clock (RLC) model.

Due to these complications, some authors employ uncorrelated relaxed clocks such as the  uncorrelated log-normal relaxed molecular clock \citep{drummond2006relaxed}, but this generates excessive rate heterogeneity in cases where clock rate changes are thought to be more punctuated, for example between HIV subtypes \citep{bletsa2019divergence}.
For a more in-depth review of various molecular clock models, see \cite{ho2014molecular}.
Here we propose an autocorrelated clock model where we place a Bayesian bridge shrinkage prior on the increment between parent and child log branch rates.
Among various shrinkage priors in the literature, the Bayesian bridge has a unique advantage in having both a collapsed spike-and-slab representation as well as a Gaussian scale-mixture form.
The first representation intuitively places large mass near zero reflecting our \textit{a priori} belief that most increments should be zero but has heavy tails that allow for estimating large rate changes in an approximately unbiased manner.
Like many other shrinkage priors, the Bayesian bridge includes a ``global scale" nuisance parameter about which learning, in the absence of prior information, typically limits the speed of inference.
\cite{polson2014bayesian} develop a framework to facilitate efficient Gibbs sampling of this nuisance parameter in a regression context and we utilize their approach here. 
On the other hand, the second representation of the Bayesian bridge as Gaussian scale-mixture is differentiable almost everywhere.
We exploit this feature and develop an efficient Hamiltonian Monte Carlo (HMC) sampler over the space of increments that employs recent work on closed form gradient representations \citep{ji2020gradients} to make our shrinkage-clock inference scalable to large trees.  
Crucial to our inference, we define recursive algorithms to compute the requisite joint gradient of the log posterior in our transformed increment space with computational complexity that scales only linearly with the number of tips in the tree.
We implement our method in BEAST \citep{suchard2018}, a popular software package for reconstructing rooted, time-measured phylogenies.
Due to our efficient inference machinery, our shrinkage-clock achieves the tractable benefits of the autocorrelated rate model and simultaneously maintains the flexibility of more punctuated local clock models.

We demonstrate the inference gains of our approach versus the RLC across 20 different simulated data sets of a 40 taxa tree.
We additionally compare the accuracy of our shrinkage-clock to the RLC by studying the adaptive radiation of rodents and other mammals, and demonstrate utility of the heavy-tailed Bayesian bridge shrinkage prior by comparing it to the more ubiquitous Laplace prior.
Finally, we deploy our shrinkage-clock to estimate the existence, location and magnitude of host-specific clock rates in surface glycoproteins of the influenza A virus.

\section{Shrinkage-based random local clocks}
\label{sec:ch4Model}
\subsection{Setup}
Consider a rooted, bifurcating tree $\phylogeny$ with $\nTaxa$ tips and $\nTaxa - 1$ internal (ancestral) nodes.
We index tips $\nodeIndexOne = 1, \ldots, \nTaxa$ and internal nodes $\nodeIndexOne = \nTaxa + 1, \ldots, 2\nTaxa - 2$.
We designate node $2\nTaxa - 1$ to be the root of the tree.
Let $\parent{\nodeIndexOne}$ denote the parent of the $\nodeIndexOne$th node and let branch length $\branchLength{\nodeIndexOne}$ connect node $\nodeIndexOne$ with its parent.

\newcommand{\siteIndex}{k}
\subsection{The relaxed clock}
Aligned molecular sequence data $\sequenceData$ evolve according to a continuous time Markov process defined by infinitesimal rate matrix $\rateMatrix$.
In our examples,
$\rateMatrix$ is a $4 \times 4$  matrix that describes the relative substitution process between nucleotides along the branches in $\phylogeny$, but in general, $\rateMatrix$ may be of larger dimension to accommodate alignments at the codon or amino acid resolution, see \cite{yang2014molecular} for reference.
Each site $\siteIndex$ of $\sequenceData$ evolves independently and identically according to $\rateMatrix$ but may have its own site-specific rate of evolution $\rateVariation$.
A priori we specify that $\expectation{\rateVariation} = 1$.
Under the relaxed clock model, the transition probability matrix for branch $\nodeIndexOne$,
\begin{equation}
\transitionMatrix_{\nodeIndexOne} = \exponential{ \scaledRate{\nodeIndexOne} \branchLength{\nodeIndexOne} \rateVariation \rateMatrix},
\end{equation}
where branch-rate multiplier $\scaledRate{\nodeIndexOne}$ is the number of expected substitutions per unit time.
To resolve identifiability issues between the height of the tree and branch-rate multipliers $\allScaledRates = \{ \scaledRate{1}, \ldots, \scaledRate{2\nTaxa - 2} \}$, we employ the rescaling proposed by \cite{drummond2010bayesian}.
 Under this transform, each branch-rate multiplier is the product of clock rate $\rate{\nodeIndexOne}$ and location parameter $\location$ scaled by the inverse of total expected substitutions per total tree time,
\begin{equation}
	\scaledRate{\nodeIndexOne} = \location
\rate{\nodeIndexOne} \frac{\sum_{\nodeIndexThree} \branchLength{\nodeIndexThree}}{\sum_{\nodeIndexThree} \rate{\nodeIndexThree}\branchLength{\nodeIndexThree}}.
\end{equation}
This results in one fewer degree of freedom since
\begin{equation}
\frac{\sum_{\nodeIndexOne} \scaledRate{\nodeIndexOne} \branchLength{\nodeIndexOne}}
{\sum_\nodeIndexOne \branchLength{\nodeIndexOne}} = \location.
\end{equation}
For heterochronous data, we estimate $\location$, but for ultrametric studies where the height of the tree is not identifiable we fix $\location = 1$.


\subsection{Autocorrelated shrinkage-clock}

\begin{figure}
	\begin{center}
		\includegraphics[width=.55\textwidth]{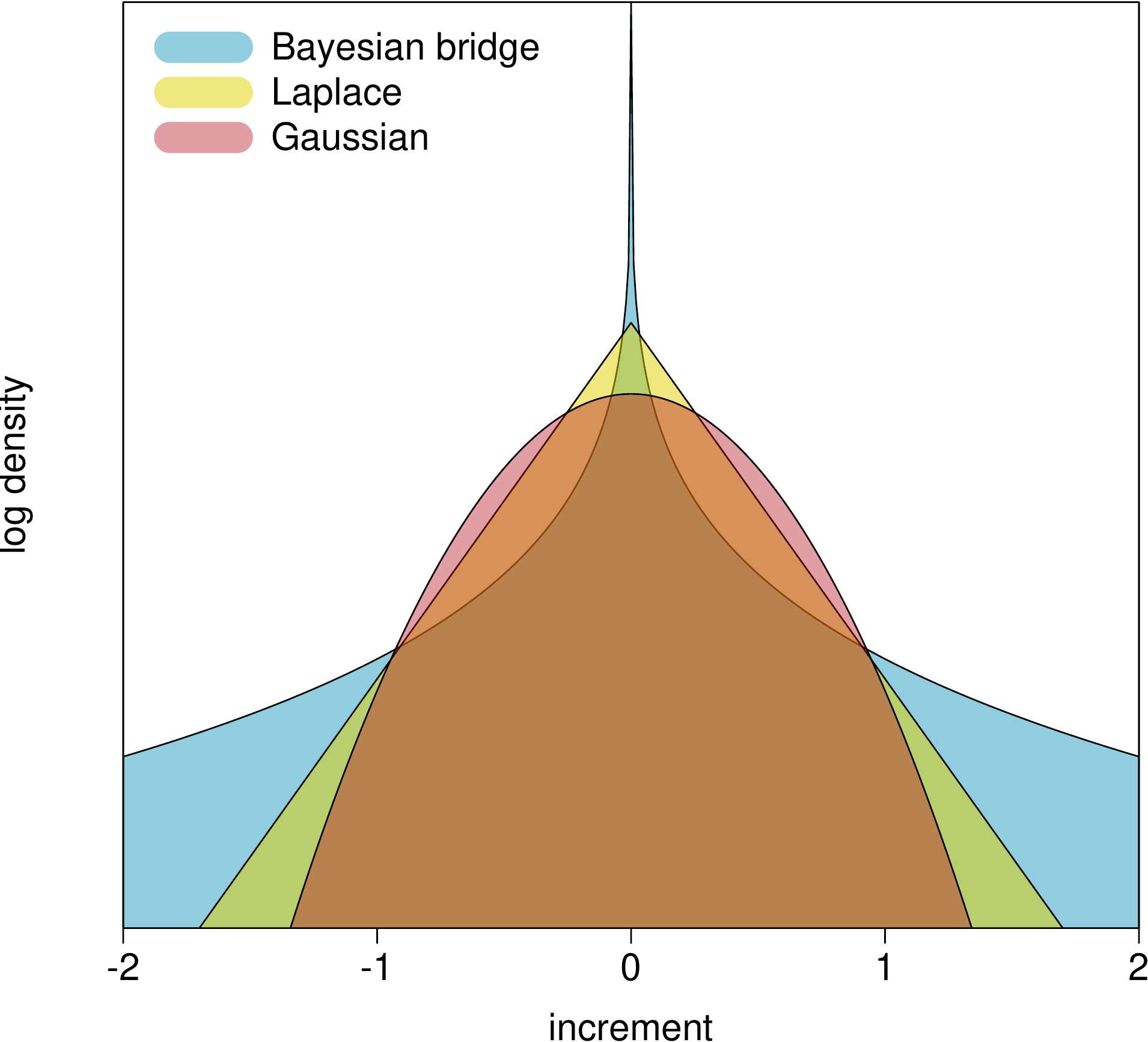}
	\end{center}
	\caption{The Bayesian bridge prior places more mass near 0 and has heavier tails compared to other common shrinkage priors.
	The Bayesian bridge reflects our a priori belief that local clocks are rare, but may arbitrarily speed-up or slow-down the rate of molecular evolution.}
	\label{fig:priorFig}
\end{figure}

We assume clock rates $\allRates = \{ \rate{1}, \ldots, \rate{2 \nTaxa - 2} \}$ are autocorrelated and model the incremental difference $\increment{\nodeIndexOne}$ between branch $\nodeIndexOne$'s clock rate and its parent lineage clock rate,
\begin{equation}
	\label{eqn:incrementTransform}
\log \rate{\nodeIndexOne} - \log \rate{\parent{\nodeIndexOne}} = \increment{\nodeIndexOne}, \ \ \text{for} \
i \in \{1, \ldots, 2\nTaxa - 2 \}
\ \text{and} \ \rate{2\nTaxa - 1} = 1.
\end{equation}
Under this parameterization, the increments $\allIncrements = \set*{\increment{1}, \ldots \increment{2\nTaxa - 2} } \in \mathbb{R}^{2 \nTaxa -2}$ are a linear transformation of $\log \allRates$.
To shrink the total number of rate changes along the tree, we let $\increment{\nodeIndexOne} \iidSim \incDistribution$ such that $\expectation{\increment{\nodeIndexOne}} = 0$.
Typically, $\incDistribution$ may follow a Gaussian \citep{thorne1998estimating} or Laplace distribution.
We choose the flexible, heavy-tailed, Bayesian bridge prior \citep{polson2014bayesian} on the increments,
\begin{equation}
\incDistribution \propto
\exp \set*{ - \left| \frac{\increment{\nodeIndexOne}}{\globalVar} \right| ^{\exponent} } ,
\end{equation}
where $\globalVar > 0$ is termed the ``global scale'' and $\exponent \in (0, 1]$ 
changes the shape of $\incDistribution$ where smaller $\exponent$ places more mass near zero.
See Figure (\ref{fig:priorFig}) for a comparison of the bridge to common shrinkage priors.
Choosing $\exponent$ to be close to $0$ forces the Bayesian bridge prior closer to best subset selection when used in a regression setting, while $\exponent = 1$ matches the Laplace prior.
In all examples, we set $\exponent = \frac{1}{4}$ to enforce slightly stronger shrinkage than the default $0.5$ employed by \cite{polson2014bayesian}. 
Since increments are independent, the joint prior is simply the product
\begin{equation}
	\label{eqn:incrementPrior}
	\cDensity{\allIncrements}{\globalVar}
	\propto
	\prod_{i=1}^{2\nTaxa - 2} \exp \set*{ - \left| \frac{\increment{\nodeIndexOne}}{\globalVar} \right| ^{\exponent} }.
\end{equation}




\section{Inference}

We follow the computationally efficient sampling approach outlined by \cite{polson2014bayesian} and view the prior on the increments as a scale mixture of normals \citep{west1987scale},
\begin{equation}
	\label{eqn:incrementDensity}
		\cDensity{\increment{\nodeIndexOne}}{\globalVar}
		= \int
\cDensity{\increment{\nodeIndexOne}}{\localVar{\nodeIndexOne}, \globalVar} \ d\localVar{\nodeIndexOne},
\end{equation}
where the local scale of branch $\nodeIndexOne$, $\localVar{\nodeIndexOne} > 0$ and draws from a one-sided stable distribution, see \cite{polson2014bayesian,nishimura2019shrinkage} for more details.
To improve convergence speed and maintain the benefits of our heavy-tailed prior, we employ the shrunken-shoulder regularization of \cite{nishimura2019shrinkage} and augment our bridge to have light tails past a reasonably large point.
Our scale mixture prior on an increment becomes
\begin{equation}
	\cDensity{\increment{\nodeIndexOne}}{\localVar{\nodeIndexOne}, \globalVar} =
\nDensity{0}{\left(\frac{1}{\slab^2} + \frac{1}{\localVar{\nodeIndexOne}^2 \globalVar^2}\right)\inverse},
\end{equation}
where slab width $\slab$ bounds the variance of increments to $\slab^2$.
In the examples to follow, we set $\slab = 2$, effecting a weakly informative, generous upper bound on clock rate changes.
Specifically, a slab width of $2$ asserts that there is at most $5 \%$ probability for $\rate{\nodeIndexOne}$ to be greater than $50 \times \rate{\parent{\nodeIndexOne}}$.
\par
We are interested in learning about the posterior,
\begin{equation}
	\label{eqn:posterior}
	\cDensity{\allRates, \globalVar, \location, \relevantParameters, \phylogeny}{\sequenceData}
	\propto
	\int
	\underbrace{
	\cDensity{\sequence}{\allRates, \location, \relevantParameters, \phylogeny}
	}_{\text{likelihood}}
	\underbrace{
	\cDensity{\allRates}{\allLocalVar, \globalVar}
	\density{\allLocalVar}
	\density{\globalVar}
	\density{\location}
	\density{\relevantParameters, \phylogeny}
	}_{\text{priors}}
	\ d \allLocalVar ,
\end{equation}
where $\allLocalVar = \{ \localVar{1}, \ldots, \localVar{2\nTaxa - 2}\}$, $\relevantParameters$ represents all relevant parameters that describe the molecular substitution model and, again, $\phylogeny$ is the phylogenetic tree.
We place a relatively uninformative, Gamma prior on $\globalVar ^{-\frac{1}{\exponent}}$ with shape $1$ and scale $2$.
Additionally, we place the CTMC conditional reference prior of \cite{ferreira2008bayesian} on the location $\location$.
We detail the priors on $\relevantParameters$ and $\phylogeny$ in each example of the sequel.

We use Markov chain Monte Carlo (MCMC) to marginalize over local scale parameters and approximate the posterior (\ref{eqn:posterior}).
Specifically, we employ a random-scan Metropolis-within-Gibbs \citep{liu2008monte,levine2006} sampling approach to update the full conditional  densities implicit in (\ref{eqn:posterior}).
Efficient sampling schemes for $\location, \relevantParameters, \phylogeny$ are well described by \cite{suchard2018}, while \cite{polson2014bayesian} outline the efficient scale-mixture approach we use to sample $\left(\globalVar, \allLocalVar\right)$.
Here, we turn our attention to sampling
\begin{equation}
	\cDensity{\allRates}{\globalVar, \location, \relevantParameters, \phylogeny, \sequenceData} \propto \int
	\cDensity{\sequence}{\allRates, \location, \relevantParameters, \phylogeny}
	\cDensity{\allRates}{\allLocalVar, \globalVar}
	\ d \allLocalVar.
\end{equation}
Since there are $2\nTaxa - 2$ correlated branch-rate multipliers, one for each branch of the tree, univariable MCMC sampling schemes for $\allRates$ scale poorly to large trees.
To remedy this difficulty, we employ Hamiltonian Monte Carlo (HMC) to sample all $\allRates$ simultaneously and with high acceptance probability.
HMC leverages the geometry of the high-dimensional branch-rate multiplier space to propose states that are farther away than traditional proposals but stay within regions of high posterior density.
HMC escapes entrapment by local extrema of the posterior by generating random momentum $\allMomentum = \{ \momentum{1}, \ldots, \momentum{2\nTaxa - 2} \}$ in each dimension where typically $\allMomentum \sim \normalDensity{\mathbf{0}}{\massMatrix}$ \citep{neal2011mcmc}.
Often mass matrix $\massMatrix=\identityMatrix{2\nTaxa-2}$ but HMC sampling may be improved by using an alternative $\massMatrix$, such as an approximation of the Hessian of the log-posterior \citep{Stan2017, zhang2011quasi}.
For further reading on HMC, see \cite{neal2011mcmc, betancourt2017conceptual}.
While HMC samplers offer more efficient posterior exploration, they require computationally expensive gradient calculations that often diminish their usefulness.
Here we exploit and extend recent work
\citep{ji2020gradients} on branch-specific clock rate gradients to facilitate fast inference of $\allRates$ under our shrinkage model.


\section{Hamiltonian Monte Carlo increment sampler}

We generate proposals in increment space, since $\allIncrements$ are uncorrelated in the prior and we transform back to rate space as described by the linear transformation in equation (\ref{eqn:incrementTransform}).
HMC sampling of the rates requires the gradient of the rate log-posterior,
\begin{equation}
\gradient{\increment{\nodeIndexThree}} \log \cDensity{\allRates}{\globalVar, \location, \relevantParameters, \phylogeny, \sequenceData} = \int
\gradient{\increment{\nodeIndexThree}}
\underbrace{
\log
\cDensity{\sequence}{\allRates, \location, \relevantParameters, \phylogeny}
}_{\condensedLikelihood}
+
\gradient{\increment{\nodeIndexThree}}
\log
\cDensity{\allRates}{\allLocalVar, \globalVar}
\ d \allLocalVar.
\label{jointGradient}
\end{equation}
To compute the gradient of the log-likelihood with respect to the increments, we first find the gradient with respect to clock rates $\allRates$,
\begin{equation}
	\gradient{\rate{\nodeIndexTwo}} \condensedLikelihood
	=
	\location
	\fullGradient{\allScaledRates} \condensedLikelihood
	\transformMatrix,
\end{equation}
where we compute all entries in $\fullGradient{\allScaledRates} \condensedLikelihood = (\gradient{\scaledRate{1}},
\ldots, \gradient{\scaledRate{2\nTaxa - 2}}) \condensedLikelihood$ with the computational $\order{\nTaxa}$  algorithm derived by \cite{ji2020gradients} and
\begin{equation}
\transformMatrix_{\nodeIndexOne \nodeIndexTwo} =
	\begin{cases}
		\frac{\sum_\nodeIndexThree \branchLength{\nodeIndexThree}}
		{\sum_{\nodeIndexThree}\rate{\nodeIndexThree} \branchLength{\nodeIndexThree}}
		-\rate{\nodeIndexOne} \branchLength{\nodeIndexOne}
		\frac{\sum_{\nodeIndexThree} \branchLength{\nodeIndexThree}}
		{( \sum_\nodeIndexThree \rate\nodeIndexThree \branchLength{\nodeIndexThree} ) ^2}  &\mbox{if } i = j \\
		-\rate{\nodeIndexOne} \branchLength{\nodeIndexTwo}
		\frac{\sum_{\nodeIndexThree} \branchLength{\nodeIndexThree}}
		{( \sum_\nodeIndexThree \rate\nodeIndexThree \branchLength{\nodeIndexThree} ) ^2} &\mbox{if } i \neq j.
	\end{cases}
\end{equation}
We complete the gradient
\begin{equation}
	\label{eqn:incrementLikelihoodGradient}
	\begin{aligned}
		\gradient{\increment{\nodeIndexThree}}
		\condensedLikelihood
&=
\sum_{\nodeIndexTwo = 1}^{2\nTaxa - 2} \gradient{\rate{\nodeIndexTwo}}
\condensedLikelihood
		\frac{d \rate{\nodeIndexTwo}}{d \increment{\nodeIndexThree}} \ \  \text{and}
		\\
		\frac{d \rate{\nodeIndexTwo}}{d \increment{\nodeIndexThree}} &=
		\begin{cases}
			\rate{\nodeIndexTwo}
			&\mbox{if i ancestral to j} \\
		0 &\mbox{otherwise,}
		\end{cases}
\end{aligned}
\end{equation}
where transformation $\frac{d \rate{\nodeIndexTwo}}{d \increment{\nodeIndexThree}}$ follows directly from equation (\ref{eqn:incrementTransform}).
To preserve the $\order{\nTaxa}$ gradient computation, we take advantage of the tree structure explicit in equation (\ref{eqn:incrementLikelihoodGradient}) and accumulate the gradient of the log-likelihood via one post-order traversal of the tree.
To begin, let $\nodeIndexOne$ and $\nodeIndexTwo$ be both daughters of node $\nodeIndexThree$ in $\phylogeny$, then
\begin{equation}
	\gradient{\increment{\nodeIndexThree}} \condensedLikelihood
	=
	\begin{cases}
	\rate{\nodeIndexThree} \times
	\gradient{\rate{\nodeIndexThree}} \left[ \condensedLikelihood \right] &\mbox{if } \nodeIndexThree \mbox{ is a tip}
	\\
	\left(
	\gradient{\increment{\nodeIndexOne}}
	+
	\gradient{\increment{\nodeIndexTwo}}
	+
	\rate{\nodeIndexThree} \times \gradient{\rate{\nodeIndexThree}}
	\right)
	\condensedLikelihood
	&\mbox{otherwise.}
	\end{cases}
\end{equation}
We next turn our attention to the gradient of the log-prior,
\begin{equation}
	\begin{aligned}
	\gradient{\increment{\nodeIndexThree}}
	\log \cDensity{\allRates}{\allLocalVar, \globalVar}
	&=
	\gradient{\increment{\nodeIndexThree}}
	\left[
	\log
	\cDensity{\allIncrements}{\allLocalVar, \globalVar}
	+
	\log
	\Bigg| \frac{d \allIncrements}{d \allRates} \Bigg|
	\right].\\
	\end{aligned}
\end{equation}
Since $\cDensity{\allIncrements}{\allLocalVar, \globalVar}$ is Gaussian, the first term gradient unwinds,
\begin{equation}
	\begin{aligned}
	\gradient{\increment{\nodeIndexThree}}
	\log
	\cDensity{\allIncrements}{\allLocalVar, \globalVar}
	&=
	\gradient{\increment{\nodeIndexThree}}
	\sum_{\nodeIndexTwo = 1}^{2\nTaxa - 2}
	\log \cDensity{\increment{\nodeIndexTwo}}{\localVar{\nodeIndexTwo}, \globalVar}
	\\
	&=
	\gradient{\increment{\nodeIndexThree}}
	\sum_{\nodeIndexTwo = 1}^{2\nTaxa - 2}
	\log \nDensity{0}{\bridgeVariance{\nodeIndexTwo}}
	\\
	&=
	- \increment{\nodeIndexThree} \fullBridgePrecision{\nodeIndexOne}.
	\end{aligned}
\end{equation}
\par
\begin{figure}[H]
	\begin{center}
		\includegraphics[width=\textwidth]{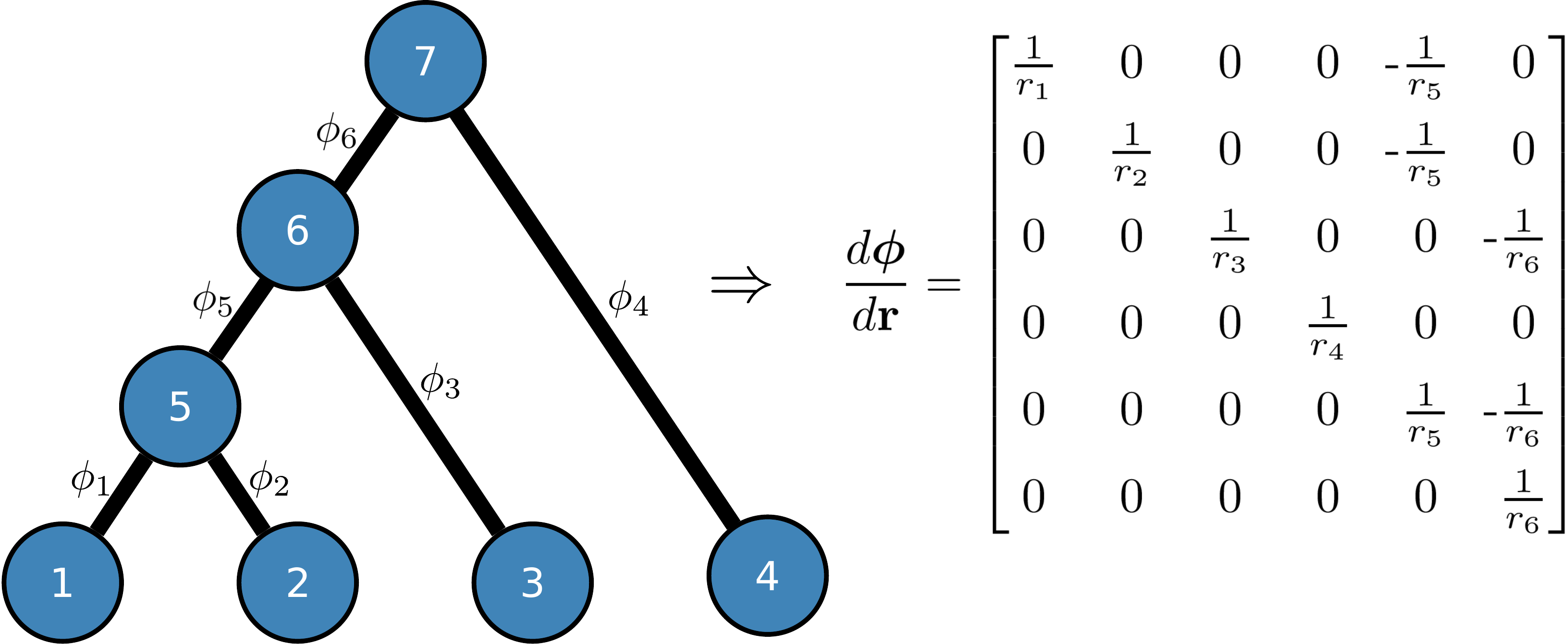}
	\end{center}
	\caption{Example tree with corresponding Jacobian matrix. Index $\nodeIndexOne < \nodeIndexTwo \implies \nodeIndexOne$ is not ancestral to $\nodeIndexTwo$ thus the Jacobian is upper-triangular and the determinant is the product of diagonal entries.}
\label{fig:bookKeeping}
\end{figure}

Numerical solutions to the second term in (\ref{jointGradient}) involve the change-of-variable Jacobian $\gradient{\increment{\nodeIndexThree}} \log \left| \frac{d \allIncrements}{d \allRates} \right|$ and appear to necessitate an $\order{\nTaxa^2}$ sparse determinant computation.
To facilitate faster computation of the transform, we index nodes of the tree such that $\nodeIndexOne < \nodeIndexTwo \implies \nodeIndexOne$ is not ancestral to $\nodeIndexTwo$.
Under this indexing, $\frac{d \allIncrements}{d \allRates}$ is an upper triangular matrix with $\frac{1}{\rate{\nodeIndexOne}}$ along its diagonal, see Figure (\ref{fig:bookKeeping}) for an example.
The gradient of the log determinant,
\begin{equation}
	\label{eqn:intermediateGradientPrior}
	\begin{aligned}
	\gradient{\increment{\nodeIndexThree}}
	\log
	\Bigg| \frac{d \allIncrements}{d \allRates} \Bigg|
	&=
	\gradient{\increment{\nodeIndexThree}}
	\log
	\prod_{\nodeIndexTwo= 1}^{2\nTaxa - 2} \frac{1}{\rate{\nodeIndexTwo}}
	\\
	&=
	-
	\gradient{\increment{\nodeIndexThree}}
	\sum_{\nodeIndexTwo = 1}^{2\nTaxa - 2}
	 \log{\rate{\nodeIndexTwo}}
	 \\
	 &=
	\underbrace{ \sum_{j} \indicatorFunction_{\left[\rate{\nodeIndexTwo} \ \text{depends on }  \increment{\nodeIndexThree} \right]} }_{\numDes{\nodeIndexThree}},
	\end{aligned}
\end{equation}
where  $\indicatorFunction$ is the indicator function and $\numDes{\nodeIndexThree}$ is the number of descendants of node $\nodeIndexThree$.
All together,
\begin{equation}
	\label{eqn:gradLogPrior}
	\begin{aligned}
	\gradient{{\increment{\nodeIndexThree}}} \log
		\cDensity{\allRates}{\allLocalVar, \globalVar}
		&=
	- \increment{\nodeIndexThree} \left(\frac{1}{\slab^2} + \frac{1}{\localVar{\nodeIndexThree}^2 \globalVar^2}\right)
	-
	\numDes{\nodeIndexThree} ,
	\end{aligned}
\end{equation}
and we accumulate $\numDes{\nodeIndexThree}$ in one recursive post-order tree traversal by observing $\numDes{\nodeIndexThree} = \numDes{\nodeIndexOne} + \numDes{\nodeIndexTwo} + 1$, where, again, $\nodeIndexOne$ and $\nodeIndexTwo$ are both daughters of $\nodeIndexThree$.

To further improve the proposals of our HMC sampler, we precondition the mass matrix $\massMatrix$ to be the current-state absolute value of the Hessian of the log-prior,
\begin{equation}
	\label{eqn:preconditioning}
	\left|
	\gradientTwo{\increment{\nodeIndexOne}}{\increment{\nodeIndexTwo}}
\log
\cDensity{\allRates}{\allLocalVar, \globalVar}
\right|
=
	\begin{cases}
		\left(\frac{1}{\slab^2} + \frac{1}{\localVar{\nodeIndexOne}^2 \globalVar^2}\right) &\mbox{if } i = j \\
0  &\mbox{otherwise.} \\
	\end{cases}
\end{equation}
This diagonal matrix weights momentum draws by prior increment precision.
Intuitively, equation (\ref{eqn:preconditioning}) improves HMC sampling by rescaling increment proposals by the variance of $\allIncrements$, allowing larger steps to be taken in dimensions with larger variance.
See \cite{neal2011mcmc} for further discussion on mass matrix transformations.

\section{Results}
\subsection{Local clocks in three nuclear genes of rodents and other mammals}
\label{sec:ex_aiv}
\begin{figure}[H]
				\begin{center}
					\includegraphics[width=\textwidth]{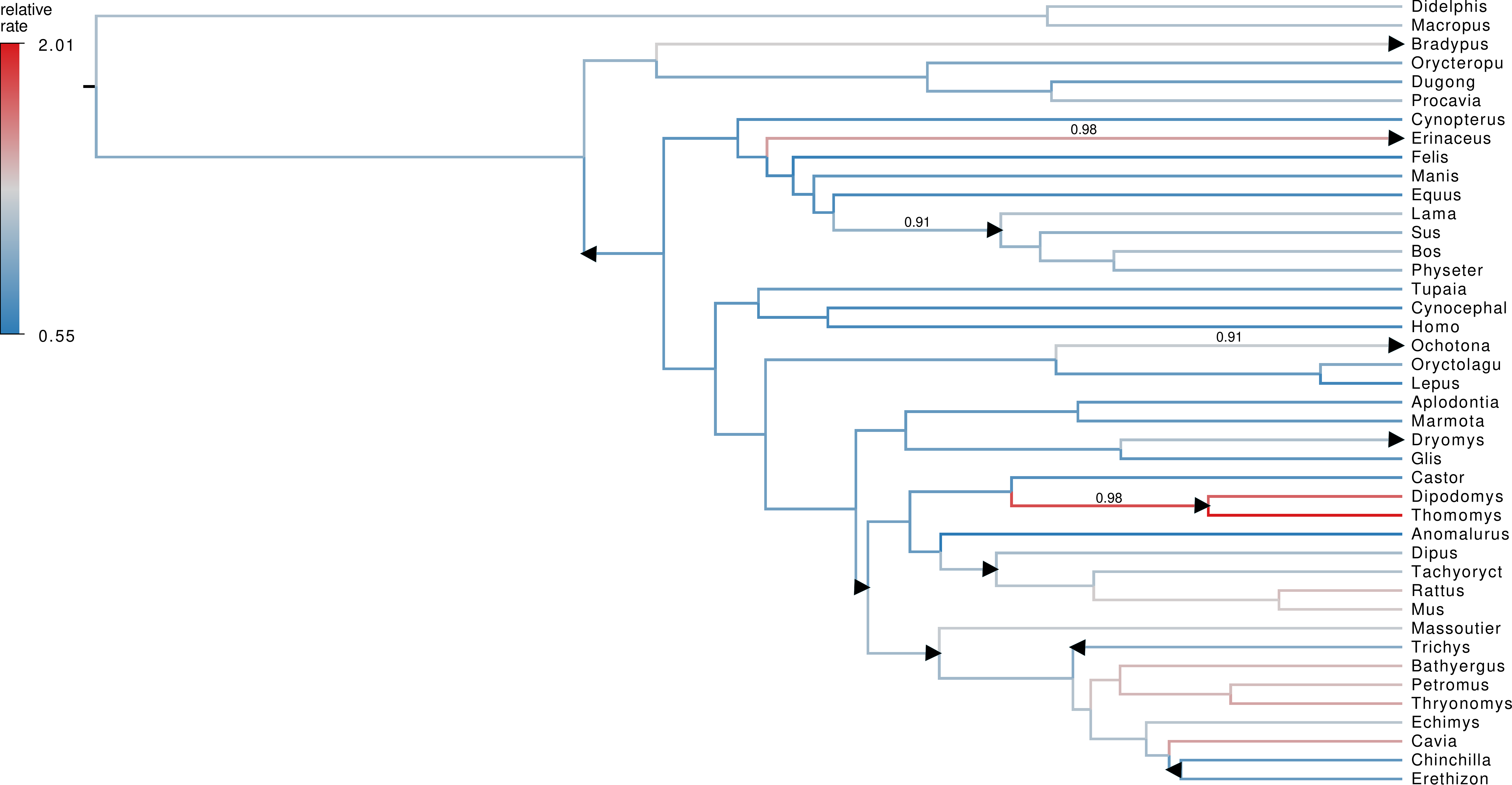}
				\end{center}
				\caption{Maximum clade credibility (MCC) tree under shrinkage-clock of mammalian and rodent radiation where branches are colored by posterior mean relative clock rates $\allRates$. If branch $\nodeIndexOne$ starts a new clock, it is labeled with the posterior probability $\increment{\nodeIndexOne}>0$. For comparison, local clocks of the random local clock (RLC) model are depicted as black triangles. Two local clocks of the RLC are excluded due to topological differences between the RLC and shrinkage-clock MCC trees.}
		\label{fig:aiv12tree}
\end{figure}

To verify the accuracy of our model, we turn to a well-studied example of adaptive radiation in mammals and rodents.
\cite{huchon2002rodent} and \cite{douzery2003local} examine the adaptive radiation of 21 rodents compared to 19 other placental mammals and two marsupial outgroups using the first two codon positions for three nuclear genes: ADRA2B, IRBP and vWF (2422 alignment sites).
\cite{douzery2003local} establish the presence of clock variability within this set of taxa and report their best fitting model contains five local clocks.
\cite{drummond2010bayesian} use the RLC model to re-examine this claim and estimate the existence of between 6 and 12 local clocks.
Here we employ our shrinkage-clock model to jointly infer the mammalian phylogeny as well as the number and location of local clocks.
Because this is an ultrametric example, $\location = 1$.
We follow the specifications of \cite{drummond2010bayesian} and \cite{douzery2003local} and use a general time reversible (GTR) substitution model with a 4 category discrete-$\Gamma$ site rate model.
We run ten separate Markov chains of our shrinkage-clock with ten different starting trees for 30M states and build a maximum clade credibility (MCC) tree from the combined results (Figure \ref{fig:aiv12tree}).
We further run 100 RLC chains with 100 different starting trees for 30M states and build a MCC tree for comparison.
Under the combinatorial parameter space of the RLC, we observe suboptimal mixing and that some chains convergence to different modes, hence our choice for combining 100 independent chains; the 10 independent chains for the shrinkage-clock simply errs on the side of caution since each independent shrinkage-clock chain converges to the same topology. 
Incidentally, the shrinkage-clock MCC topology differs from the RLC MCC in two places.
First, \textit{Bradypus} attaches to one of two neighbor internal branches deep in the tree.
Second, \textit{Anomalurus} is more closely related to the \textit{Dipus} than \textit{Castor} under the shrinkage-clock.
This second difference highlights the well-known difficulty in \textit{Anomalurus} placement \citep{horner2007phylogenetic}.  
Indeed, \cite{blanga2009rodent} find \textit{Anomalurus} sits between \textit{Dipus} and \textit{Castor} in an analysis of six nuclear genes.
The posterior probabilities of \textit{Bradypus} and \textit{Anomalurus} parent branches under the shrinkage-clock are almost equal (0.64 and 0.49 respectively).

We estimate the existence of four local clocks where we define a local clock on branch $\nodeIndexOne$ if the posterior odds $\increment{\nodeIndexOne} > 0$ is greater than $10$ or less than $\frac{1}{10}$.
The posterior odds here is equivalent to a Bayes factor since an increment is equally likely to be positive or negative under the prior.
Furthermore, a Bayes factor greater than $10$ is suggestive of ``strong evidence'' against an alternative hypothesis \citep{kass1995bayes}.
In the approach, we do not make assumptions about the magnitude of local clocks on a tree and instead use posterior probability of increment sign to define a clock.

To illustrate the benefits of using a heavy-tailed prior on the increments, we further compare the performance of our Bayesian bridge prior to the more usual Laplace prior for shrinkage (see Figure \ref{fig:priorFig}).
We again fit our shrinkage-clock as described above but remove the slab and fix $\exponent = 1$, thus placing a Laplace prior on each increment.
We find the posterior mean of increment variance under both the Laplace and Bridge priors is 0.057 with 95\% high posterior density (HPD) intervals $\left( 0.036, 0.089 \right)$ and $\left( 0.035, 0.093 \right)$ respectively.
Despite having very similar variance, we find the posterior mean of the absolute maximum increment is 0.84 $\left(0.57, 1.22 \right)$ and 1.01 $\left( 0.70, 1.54 \right)$ under the Laplace and Bridge priors respectively.
This evidences induced smoothing of the clock rates under the exponential tails of a Laplace prior, that on average shrinks the largest increment by approximately 20\%.

\subsection{Simulation study}
We compare the scalability of our shrinkage-clock to the RLC under a simulated example.
We generate 1000 character nucleotide sequences from a fixed 40 tip tree 20 times.
In each simulation, there are 4 distinct lineages (A, B, C, D) of 10 taxa each.
Time to most recent common ancestor (TMRCA) for each lineage is 40 years and tree height is 80 years.
Lineages B, C and D evolve with a relative clock rate of $1.0$ while the MRCA of lineage A starts a new clock with relative rate $2.0$.

\begin{figure}[H]
	\begin{center}
		\includegraphics[width=\textwidth]{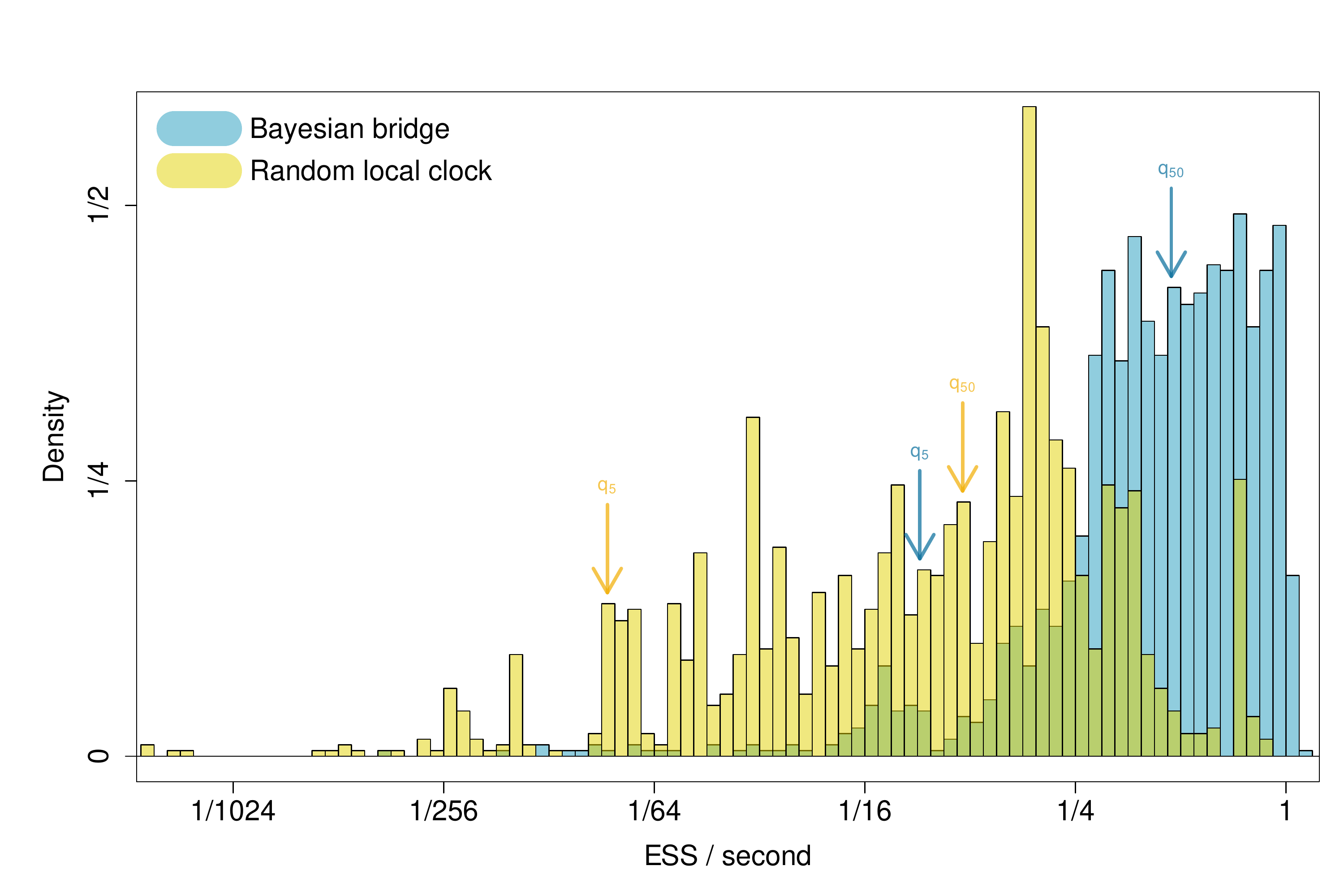}
	\end{center}
	\caption{Effective sample size of branch-specific clock rates per second of BEAST runtime under the shrinkage-clock and RLC during a full joint phylogenetic analysis.}
	\label{fig:ess}
\end{figure}

We compare the accumulation of effective sample size (ESS) per unit time of branch-specific clock rates under both our Bayesian bridge shrinkage-clock and the RLC while simultaneously inferring the phylogeny.
ESS approximates the number of independent samples from a Markov chain and we use this metric to evaluate how well each inference procedure explores clock rate space.
We report the results across all 20 simulated datasets in Figure (\ref{fig:ess}).
The median ESS/second is 0.49 and 0.13 under the shrinkage-clock and the RLC respectively, exhibiting a 3.8-fold speed increase.
Additionally, the ``least well" explored clock rate across all simulations accumulated $3.5 \times 10^{-3}$ and $4.9 \times 10^{-4}$ ESS/second under the shrinkage-clock and RLC respectively, a 7.1-fold speed-up, for these relatively small taxon-count examples.

We further report here that inference under the shrinkage-clock without preconditioning the mass matrix results in a minimum and median ESS/second of $2.3 \times 10^{-4}$ and $1.5 \times 10^{-3}$ respectively.
Overall, preconditioning the mass matrix improves ESS/second $15\times$ for the worst-explored clock rate under the shrinkage-clock.

Averaged across all twenty simulations, the Bayes factor for the true clock rate change under our shrinkage clock is $5.25$ while the second most likely clock rate has Bayes factor support of $0.56$.
Comparably, averaged over all runs, the Bayes factor of the one true clock under the RLC is $3.51$. 
Additionally, the true relative clock-rate for the `A' clade is $2.0$ and we estimate $1.51$ $\left( 0.90, 2.31 \right)$  and $1.49$ $\left( 0.98, 2.36 \right)$ under the shrinkage-clock and RLC respectively.

\subsection{Influenza A virus}
We further demonstrate the scalability and utility of our shrinkage-clock model by examining the evolution of two major influenza A virus (IVA) surface glycoprotein subtypes: hemagglutinin (HA) H7 and neuraminidase (NA) N7.
Both hemagglutinin and neuraminidase protein mutations impact IVA's epitope and allow IVA to escape adaptive immune responses \citep{mcauley2019influenza, wilson1990structural}.
\cite{worobey2014synchronized} find divergence time estimation is sensitive to molecular clock model specification.
To consistently estimate divergence times, \cite{worobey2014synchronized} allow the clock rates of various glycoprotein subtypes to vary only between viral hosts and find H7 and N7 each evolve slower in equine hosts than avian hosts.
We re-examine this claim with our more general shrinkage-clock model that does not assume the existence of host-dependent clock rates.
Specifically, we re-analyze 146 complete gene (1716 nt) sequences 
 of H7 and 92 complete gene (1416 nt) sequences of N7. 
In each case we follow the model specifications of \cite{worobey2014synchronized} and employ a GTR substitution model with 4 category discrete-$\Gamma$ site rate model.
We depart from their example, however, in our use of tree prior. We employ a Bayesian skygrid prior \citep{gill2013improving} with 50 population size bins and a cutoff of 200 years instead of using the skyride prior \citep{minin2008smooth}.
While the number of taxa here is only approximately double or triple the number in the previous simulation study, the set of all possible local clocks under the RLC grows exponentially with the number of taxa, (31 to 63 orders of magnitude), challenging clock rate inference under the RLC.
\begin{figure}[H]
	\begin{center}
		\includegraphics[width=\textwidth]{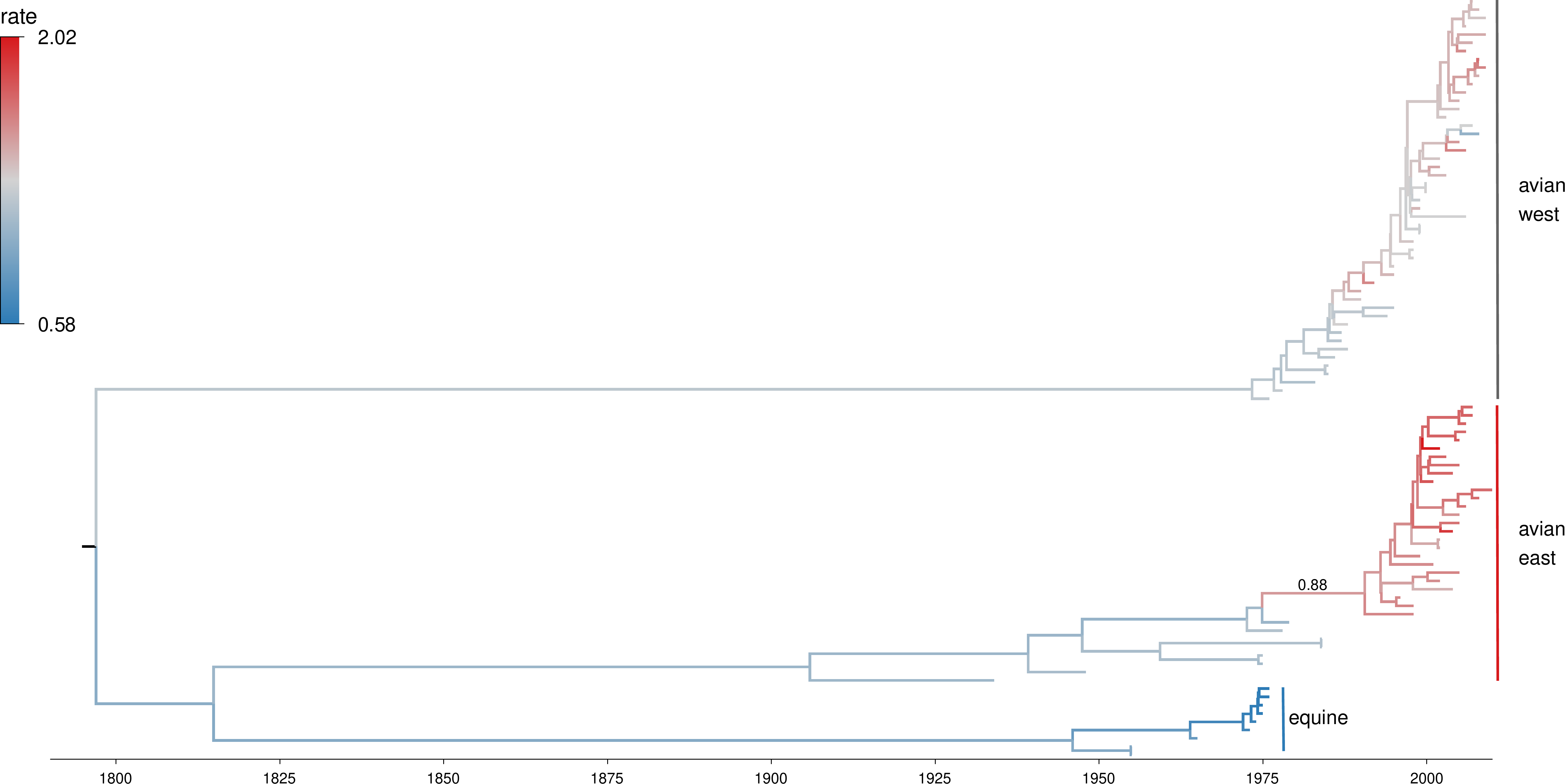}
	\end{center}
	\caption{Maximum clade credibility tree for influenza A's neuraminidase subtype N7. Branches are colored by posterior clock rates. The most probable local clock is reported and labeled with posterior probability that $\increment{\nodeIndexOne} >0$. The second most probable clock starts a sub-clade of the equine lineage and has a Bayes factor $\increment{\nodeIndexOne} >0$ of 0.351.}
	\label{fig:N7_MCC}
\end{figure}


\par
\begin{figure}[H]
	\begin{center}
		\includegraphics[width=\textwidth]{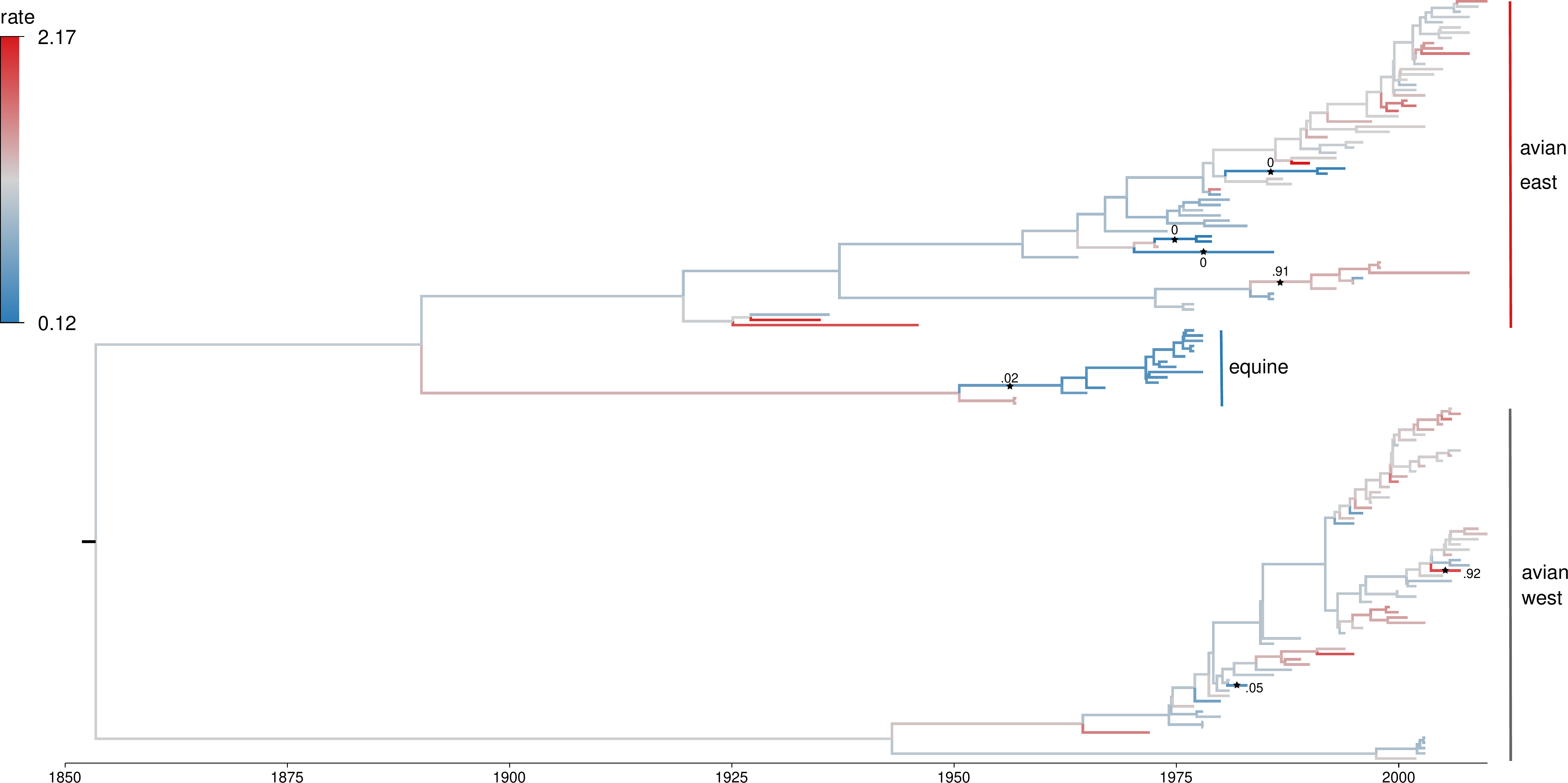}
	\end{center}
	\caption{Maximum clade credibility tree for influenza A's hemagglutinin subtype H7. Branches are colored by posterior clock rates. Local clocks are labeled with a star and the posterior probability $\increment{\nodeIndexOne} > 0$.}
	\label{fig:H7_MCC}
\end{figure}

We find no sharp local clocks exist with Bayes factor $>10$ or $<\frac{1}{10}$ on the NA N7 tree under our shrinkage-clock model but do see evidence for rate heterogeneity.
Incidentally, the most likely clock occurs on the branch that begins the Eastern avian clade of the NA N7 tree (Figure \ref{fig:N7_MCC}).
The second most probable clock is found in a subclade of the equine lineage.
On the other hand, we estimate the existence of seven local clocks on the HA H7 tree and report these in Figure (\ref{fig:H7_MCC}).
Overall, the mean posterior clock rate for NA N7 is lower than HA H7. 
We report the posterior mean and 95\% HPD intervals of $\location$ are $2.7\ \{2.1 - 3.3 \}  \times 10^{-3}$ and $3.3\ \{ 2.8 - 3.9  \} \times 10^{-3}$ for NA N7 and HA H7 respectively.
Furthermore, under our shrinkage-clock the posterior mean root heights and $95\%$ HPD intervals of the N7 and H7 trees in absolute time are 1798 (1733-1855) and 1853 (1808 - 1897) respectively.

\section{Discussion}
Previous heritable clock models either scale poorly to large trees or excessively shrink clock rates.
We develop a robust auto-correlated heritable clock model to overcome these challenges without specifying \textit{a priori} the number and location of local clocks on a tree.
Crucially, we model the incremental difference between log clock rates on the tree as drawing from a Bayesian bridge prior that shrinks most changes to approximately 0 unless the data warrant otherwise.
To facilitate scalability, we employ HMC to generate proposals in the independent increment space and derive recursive post-order algorithms to compute the gradient and its requisite transforms.
Our recursive algorithms achieve $\order{\nTaxa}$ computational speed, signifying that they will continue to work well as $\nTaxa$ grows large.
We further improve the speed of our HMC sampler by preconditioning the mass matrix with the Hessian of the log-prior.

In our examination of the adaptive radiation of rodents and other mammals, (see section \ref{sec:ex_aiv}), our shrinkage-clock recovers the location of four local-clocks estimated under the RLC.
This finding is similar to the initial estimate of five clocks reported by \cite{douzery2003local}.
We choose a Bayes factor of $10$ to classify clocks, but shrinkage-clock users may wish to adjust this threshold to increase or decrease clock rate sensitivity.
Comparing the statistical properties of different clock-classification schemes remains an important avenue  for future work.

We apply our shrinkage-clock model to re-examine the number of local clocks present in IVA surface glycoproteins NA N7 and HA H7 across equine and avian hosts.
We confirm the equine slowdown reported by \cite{worobey2014synchronized} but interestingly find that the N7 tree shows marked rate variation (Figure \ref{fig:N7_MCC}) between western and eastern hemisphere avian influenza lineages, however, this rate variation is not supported by the Bayes factor cutoff. 
Root height estimates vary from \cite{worobey2014synchronized} but this may be in part due to the different tree priors.
Additionally, we find 7 local clocks under our shrinkage-clock across the H7 tree.
Since three of these clocks belong to edges of tip nodes, this may reflect incomplete sampling or sequencing error.
Despite inferring six more clocks than the host-specific model, the posterior mean estimate of root height under our shrinkage-clock is within five years of previous estimates \citep{worobey2014synchronized}.

Our shrinkage-clock accumulates ESS/second of the worst explored clock rate $7.1 \times$ faster than the RLC  across 78 branches of 20 simulated data sets.
If ESS is used as stopping criteria for phylogenetic reconstruction, this could save users up to 85\% of BEAST runtime.
As the bridge exponent $\exponent$ approaches $0$, the bridge prior density is more peaked near zero resulting in sharper increment shrinkage and thus better distinguishable local clocks.
\cite{nishimura2019shrinkage} examine multiple $\exponent$ and report closer to optimal coverage for smaller $\exponent$ but with increasing computational cost due to mixing.
For this reason, shrinkage-clock users may find it useful to adjust $\exponent$ depending on desired clock-rate coverage or to tackle even larger tree studies.
We make all BEAST XML files used in this work publicly available at \href{https://github.com/suchard-group/shrinking_clocks}{https://github.com/suchard-group/shrinking\_clocks}.

\section{Acknowledgements}
This work was supported through National Institutes of Health grants R01AI153044 and U19AI135995, European Research Council under the European Union's Horizon 2020 research and innovation programme (grant agreement no. 725422-ReservoirDOCS) and the Wellcome Trust project 206298/Z/17/Z (ARTICNetwork).
PL acknowledges support by the Research Foundation -- Flanders (`Fonds voor Wetenschappelijk Onderzoek -- Vlaanderen', G066215N, G0D5117N and G0B9317N).

\bibliographystyle{chicago}

\end{document}